\documentclass[conference,a4paper]{IEEEtran}
\usepackage{xcolor}
\usepackage{balance}
\usepackage{newtxmath}
\ifCLASSINFOpdf
  \usepackage[pdftex]{graphicx}
\else
  \usepackage[dvips]{graphicx}
\fi
\ifCLASSOPTIONcompsoc
 \usepackage[caption=false,font=normalsize,labelfont=sf,textfont=sf]{subfig}
\else
 \usepackage[caption=false,font=footnotesize]{subfig}
\fi
\usepackage{bm}
\usepackage{xcolor}
\usepackage{mathtools}
\usepackage{cite}
\usepackage{csquotes}
\usepackage{siunitx}
\newcommand*{\mj}   {\mathrm{j}}
\newcommand*{\me}   {\mathrm{e}}

\renewcommand*{\vec}{\bm}

\newcommand{\mat}[1]{\mathbf{#1}}

\newcommand{\norm}[1]{\left\lVert #1 \right\rVert}

\usepackage{pgfplots}
\pgfplotsset{compat=newest}
\usepackage{tikz}
\usetikzlibrary{external}
\tikzset{external/up to date check={md5}}
\tikzset{external/mode=convert with system call} 

\usetikzlibrary{plotmarks}
\newlength\fh
\newlength\fw

\usepackage{subfig}

\usepackage{placeins}

\begin{document}
\title{Linear Phase Retrieval for Near-Field Measurements with Locally Known Phase Relations}
\author{\IEEEauthorblockN{
		Alexander Paulus\IEEEauthorrefmark{1},
		Jonas Kornprobst\IEEEauthorrefmark{1},
		Josef Knapp\IEEEauthorrefmark{1}, and
		Thomas F. Eibert\IEEEauthorrefmark{1}
	}%
	\IEEEauthorblockA{\IEEEauthorrefmark{1}%
		Chair of High-Frequency Engineering, Department of Electrical and Computer Engineering, \\ Technical University of Munich, Munich, Germany, a.paulus@tum.de}}

\maketitle

\begin{abstract}
A linear and thus convex phase retrieval algorithm for the application in phaseless near-field far-field transformations is presented.
The formulation exploits locally known phase relations among sets of measurement samples, which can in practice be acquired with multi-channel receivers. 
Due to the linearity of the formulation, a reliable phaseless transformation is achieved, which completely avoids the problem of local minima\,---\,the Achilles heel of most existing phase retrieval techniques. 
Furthermore, the necessary number of measurements are kept close to that of fully-coherent antenna measurements. 
Comparisons with an already existing approach exploiting local phase relations demonstrate the accuracy and reliability for synthetic data.
\end{abstract}

\vskip0.5\baselineskip

\section{Introduction}
Over the past decades, near-field (NF) measurements with a successive NF far-field (FF) transformation (NFFFT) have established as a low-cost alternative to measurements in FF or compact ranges.
In order to further reduce the cost and to adapt to the challenges of next generation mobile communication systems, phaseless transformation algorithms are intensively investigated~\cite{Paulus.2017b, Paulus.2017,Schmidt.2009, Yaccarino.1999,Paulus.2018,Pierri.1999,Schnattinger.2014,Bucci.1999,Bangun.2019,Isernia.1996,Berlt.2020,Sanchez.2020,Moretta.2019}. Supported by the comprehensive research in the field of phase retrieval, e.g., see~\cite{Gerchberg.1972,Fienup.1982,Candes.2015, Candes.2013,Waldspurger.2015,Netrapalli.2015}, the antenna community has proposed various approaches for obtaining the FF behavior from NF measurements without reliable phase information, as required by common NFFFTs~\cite{Costanzo.2007,Bucci.1991,Schmidt.2009b,Petre.1992,Jensen.1975,Hansen.1988}. Recent phaseless NF antenna measurements rely on holography~\cite{Gabor.1949} with a fixed and known reference antenna~\cite{Berlt.2020,Sanchez.2020,Laviada.2013,LaviadaMartinez.2014,Castaldi.2000}, on measurements on multiple surfaces~\cite{ Schmidt.2009, Yaccarino.1999,Isernia.1996,LasHeras.2020,Moretta.2019,Yaccarino.1997,Fuchs.2020}, apply specialized probe antennas~\cite{Costanzo.2001b,Costanzo.2005,Costanzo.2008,Paulus.2017b,Paulus.2017,Sanchez.2020} or utilize multi-frequency information~\cite{Paulus.2020,Knapp.2020}. The holographic approach is often limited by the dynamic range of the measurement setup and suffers from practical issues related to the positioning of the reference antenna without obstructing the measurement probe or antenna under test (AUT) at any time. Most of the approaches utilizing special probe antennas aim at forming chains of locally coherent measurements, such that phase relations between measured signals at distant locations can explicitly be established~\cite{Costanzo.2001b}. However, error propagation makes this approach sensitive to measurement inaccuracies. Instead of explicitly reconstructing the phase, some attempts have been made to exploit the additional information provided by the special probe antennas and by measurements on multiple surfaces with iterative phase retrieval algorithms, e.g.,~\cite{Yaccarino.1999,Paulus.2017b}. The resulting transformation results have \textit{often} been observed to be superior to the case when relying on measurements with common probe antennas and on a single surface. However, it should be noted that due to the rather complicated nature of the phase retrieval problem, no practical guidelines or convergence promises for real-world applications exist. As such, the techniques reported in literature \textit{may improve} the suitability of measurement setups for phaseless NFFFTs, however, there exists no algorithm that can reliably perform a transformation with that kind of structured data. 

Recently, a possible foundation for reliable phase retrieval has been reported~\cite{Kornprobst.07.02.2020}, when partially coherent observations, e.g., in the form of local phase relations, are available.
In the present paper, in contrast to~\cite{Kornprobst.07.02.2020}, we present a formulation in terms of mixed unknowns, corresponding to source coefficients and phase terms. In this way, the problem formulation features more flexibility while keeping the computational costs low. The presented method exploits partial coherence in the measured data while preserving linearity, and thus convexity. The approach perfectly matches the problem arising from phaseless NF antenna measurements when a multi-channel receiver with multiple probe antennas is used. No synchronization between the AUT and the probes is required, and no coherence among sequentially acquired samples is assumed. However, due to the receiver architecture, local coherence between the channels of the receiver at each respective measurement position is given. This can either be achieved with vectorial measurement hardware, or with adequate scalar equipment, however, requiring a larger number of channels. Note that the formulation can also be applied to (incomplete) holographic measurements with a reference antenna. Furthermore, the need for a fixed reference antenna is removed, diminishing the problem of limited dynamic range and mutual shadowing of probe, AUT, and reference antenna.

We start by briefly revising the common NFFFT source reconstruction problem and discuss the model of partially coherent observations. For the purpose of comparability, we recapitulate an existing formulation for exploiting partially coherent measurements, followed by the novel linear phase retrieval formulation. Results for simulation data are discussed.

\section{Phase Retrieval With Partial Coherence}
For the sake of simplicity and without loss of generality, we consider NF antenna measurement setups where the unknown AUT is transmitting and the field is sampled by one or multiple probe antennas in receive mode. The AUT is represented by a vector of complex-valued source coefficients $\vec z\in\mathbb{C}^{n}$, the probe antennas acquire the complex-valued signals $\vec b\in\mathbb{C}^{m}$ and the solution of the relation 
\begin{align}
\mat A \vec z &= \vec b\label{eq:phase}
\end{align}
represents the main task of an NFFFT with full phase information. 
In the case of phaseless measurements, only $\left|\vec b\right|$ is known and~\eqref{eq:phase} deteriorates into a nonlinear system of equations. In both cases, the desired fields, e.g., the FF, can be computed from $\vec z$, once it has been determined.

Let us now assume that we have acquired three sets of NF measurements, $\vec b_1\in\mathbb{C}^{m_1}$, $\vec b_2\in\mathbb{C}^{m_2}$, and $\vec b_3\in\mathbb{C}^{m_3}$. Without loss of generality, assume that $\vec b_1$ has been acquired with some random phase, but the second and third set of samples $\vec b_{2/3}$ have been measured \textit{simultaneously} by a two-channel receiver equipped with two probe antennas. Consequently, both sets feature the same number of entries, $m_2 = m_3$, and there is a known phase relation between the $k$th entry in $\vec b_2$ and $\vec b_3$ for $k\in\left\{1,...,m_2\right\}$.

\subsection{Interferometry via Linear Combinations}
\label{sec:interfero}
It is well known that the phase difference between two complex scalars $a_1$ and $a_2$ can be computed analytically from four magnitude measurements of the linear combinations~\cite{Costanzo.2001b}
\begin{align}
o_1 &= \left|a_1\right|^2, \hspace*{0.5cm}o_3 = \left|a_1+a_2\right|^2, \nonumber\\
o_2 &= \left|a_2\right|^2, \hspace*{0.5cm}o_4 = \left|a_1+\mj a_2\right|^2
\end{align}
via
\begin{align}
\angle\left(a_1 \overline{a_2}\right)&= \text{atan}\left[\dfrac{o_4-o_1-o_2}{o_3-o_1-o_2}\right]\label{eq:exactPR_interfero},
\end{align}
where $\overline{a_2}$ denotes the complex conjugate of $a_2$.
Coming back to our example from above, we can thus formulate the task of a phaseless NFFFT with knowledge of the partial coherence between $\vec b_2$ and $\vec b_3$ as
\begin{align}
\left|\begin{bmatrix}
\mat A_1 \\ \mat A_2 \\ \mat A_3 \\ \mat A_2+\mat A_3 \\ \mat A_2+\mj\mat A_3
\end{bmatrix}\vec z \,\right| &= \left|\begin{bmatrix}
\vec b_1 \\ \vec b_2 \\ \vec b_3 \\ \vec b_2+\vec b_3 \\ \vec b_2+\mj\vec b_3
\end{bmatrix}\right|, \label{eq:PC_LC_implementation}
\end{align}
with elementwise magnitude operations on both sides of the equation.
Note that when neglecting the added rows of linear combinations in the forward operator and the right-hand-side, the common, fully-incoherent problem is obtained. In terms of the notation in~\eqref{eq:PC_LC_implementation}, the knowledge of the phase differences thus directly translates into additional measurement rows. The overall structure remains the same.

A simple approach of exploiting partial coherence is thus to solve~\eqref{eq:PC_LC_implementation} with a phase retrieval algorithm of your choice. As reported in~\cite{Paulus.2017,Paulus.2017b}, the additional rows related to the phase differences increase the chance of a successful transformation. Still, success is not guaranteed and only limited theoretical 
insight into the behavior of the formulation is available~\cite{Knapp.2019}.

\subsection{Linear Formulation for Partial Coherence}
Here, we follow an idea presented in~\cite{Kornprobst.07.02.2020} and derive a flexible formulation with low computational cost. 
In order to illustrate the new approach, we consider an example with $m_1 = 1$, $m_2 = 2 = m_3$, and
\begin{align}
\vec b_1 = \left|b_{11}\right|\me^{\,\mj \varphi_{11}},\hspace*{0.10cm}
\vec b_2  = \begin{bmatrix}
\left|b_{21}\right|\me^{\,\mj \varphi_{21}} \\ \left|b_{22}\right|\me^{\,\mj\varphi_{22}}
\end{bmatrix},
\hspace*{0.10cm}
\vec b_3 = \begin{bmatrix}
\left|b_{31}\right|\me^{\,\mj \varphi_{31}} \\ \left|b_{32}\right|\me^{\,\mj\varphi_{32}}
\end{bmatrix}.\label{eq:example}
\end{align}

We can write
\begin{align}
&\mat A\vec z = \vec b = \begin{bmatrix}
\vec b_1^{\text{T}} & \vec b_2^{\text{T}} & \vec b_3^{\text{T}} 
\end{bmatrix}^{\text{T}} \nonumber\\&= \begin{bmatrix}
\left|b_{11}\right|   &    0                   & 0                     \\
0                     &  \left|b_{21}\right|   & 0                     \\
0                     &  0                     & \left|b_{22}\right|   \\
0                     &  \left|b_{31}\right|\me^{\,\mj \left(\varphi_{31} - \varphi_{21}\right) }   & 0                     \\
0                     &  0                     & \left|b_{32}\right| \me^{\,\mj \left(\varphi_{32} - \varphi_{22}\right) }  \\
\end{bmatrix}\begin{bmatrix}
\me^{\,\mj \varphi_{11}} \\ \me^{\,\mj \varphi_{21}} \\\me^{\,\mj \varphi_{22}}
\end{bmatrix}\nonumber \\
&= \text{diag}\left(\left|\vec b\right|\right) 
\begin{bmatrix}
1   &    0                   & 0                     \\
0                     &  1   & 0                     \\
0                     &  0                     & 1   \\
0                     &  \me^{\,\mj \left(\varphi_{31} - \varphi_{21}\right) }   & 0                     \\
0                     &  0                     &  \me^{\,\mj \left(\varphi_{32} - \varphi_{22}\right) }  \\
\end{bmatrix} \vec \psi\nonumber \\
&= \mat B \mat C \vec \psi,
\end{align}
leading to the linear formulation
\begin{align}
\left(\mat A \mat P_1 - \mat B \mat C  \mat P_2\right) \tilde{\vec z} &= \vec 0,\hspace*{0.5cm}\tilde{\vec z} = \begin{bmatrix}
\vec z^{\text{T}} & \vec \psi^{\text{T}}
\end{bmatrix}^{\text{T}}\in\mathbb{C}^{n+q} \label{eq:MAIN}
\end{align}
with the stacked vector of unknowns $\tilde{\vec z}$ consisting of the unknown AUT representation $\vec z$ and a vector $\vec \psi$ of all remaining unknown phase terms. The matrices $\mat P_{1/2}$ in~\eqref{eq:MAIN} are defined via unit and zero matrices $\mat I$ and $\mat 0$, respectively, as
\begin{align}
\mat P_1 \tilde{\vec z} &= \begin{bmatrix}
\mat I & \mat 0
\end{bmatrix}\tilde{\vec z} = \vec z \in\mathbb{C}^{n}\\
\mat P_2 \tilde{\vec z} &= \begin{bmatrix}
\mat 0 & \mat I
\end{bmatrix}\tilde{\vec z} = \vec \psi\in\mathbb{C}^{q},
\end{align}
and extract the respective parts from the stacked vector of unknowns. The diagonal matrix $\mat B\in\mathbb{R}^{m\times m}$ contains all measured magnitudes while the actual information about partial coherence is inside the sparse matrix $\mat C \in\mathbb{C}^{m\times q}$, containing all known phase differences. 
The more phase differences are added, the less independent phase terms remain unknown and the fewer columns in $\mat C$ exist.
Note that for~\eqref{eq:MAIN} to yield the correct solution, we have to enforce that $\left| \psi_k\right| = 1$, $\forall k \in \left\{1,...,q\right\}$. However, this condition is nonlinear and leads to a nonconvex problem. Still, we find an approximate and linear formulation by writing
\begin{align}
\left(\mat A \mat P_1 -\mat B \mat C \mat P_2\right)\tilde{\vec z} &= \vec 0 \label{eq:linear_main}\\
\nonumber &\hspace*{-1.8cm}\text{s.t.}\,\,\, \psi_s = 1, 
\end{align}
which only enforces the magnitude constraint on the $s$th phase unknown and fixes its phase to zero. In this way, a linear and thus convex phase retrieval formulation for the case of partially coherent observations is obtained. Note that, in contrast to (18) in~\cite{Kornprobst.07.02.2020}, the here presented notation is more flexible and allows for an arbitrary combination of incoherent and coherent measurements. Also, the large computational effort per matrix-vector product of (21) in~\cite{Kornprobst.07.02.2020}, requiring a pseudo-inverse, is avoided. As a downside of~\eqref{eq:linear_main}, more unknowns are introduced.

\section{Transformation Results}
\subsection{Randomly Generated Data}
As a first indicator for the performance of the formulation in~\eqref{eq:linear_main}, synthetic data randomly taken from a complex-valued normal distribution was generated and processed. The reference formulation in~\eqref{eq:PC_LC_implementation} was solved via the phase retrieval algorithm described in~\cite{Paulus.2017b}, which essentially is a nonconvex optimization-based phase retrieval algorithm similar to the widely known Wirtinger flow~\cite{Candes.2015}. The algorithm features an iterative line search algorithm according to~\cite{Nocedal.2006} and computes descent directions via the L-BFGS method~\cite{Liu.1989}. Due to its nonconvexity, the formulation heavily depends on the initial guess, which was computed via $\num{40}$ power iterations of a spectral method~\cite{Candes.2015}. While~\eqref{eq:linear_main} was solved via a direct inversion in Matlab~\cite{MATLAB.2019b}, at most $\num{2000}$ iterations were allowed for the iterative solution of~\eqref{eq:PC_LC_implementation}.

\begin{figure}[t]
	\centering
	\includegraphics{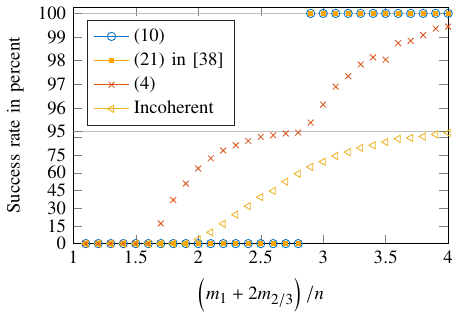}
	\caption{Comparison of the formulation in~\eqref{eq:PC_LC_implementation}, the proposed formulation in~\eqref{eq:linear_main}, formulation (21) in~\cite{Kornprobst.07.02.2020}, and~\eqref{eq:PC_LC_implementation} without the linear combinations (incoherent) for complex-valued normally distributed data. The problem dimensions were picked as $n = 20$, $m_1 = n$, and the number of partially coherent measurements $m_{2/3}$ was swept from $\num{0.05}n$ to $1.5n$. For each sampling ratio, $\num{5000}$ randomly generated measurement matrices, solution vectors, and the corresponding measurement vectors were considered.}
	\label{fig:synth_result}
\end{figure}

The model with three sets of measurements introduced in Section~\ref{sec:interfero} was utilized, with $n = 20$ and $m_1 = n$. The number of measurements with local phase relations $m_{2/3}$ was varied within the range of $\num{0.05}n$ to $1.5n$ to account for a variable contribution of phase information.
Figure~\ref{fig:synth_result} depicts the chance of a successful phase retrieval, where success was declared once a relative complex NF deviation
\begin{align}
\varepsilon_{\text{c}} &= 20\,\text{log}_{10}\left(\dfrac{\norm{\mat A \vec z-\vec b}}{\norm{\vec b}}\right)\label{eq:cNF}
\end{align} 
below $\SI{-90}{\dB}$ was achieved. Note that global phase shifts between $\mat A\vec z$ and $\vec b$ are ignored in the evaluation of~\eqref{eq:cNF} since phase retrieval is always only possible up to an arbitrary global phase shift. 
For real-world measurements, only the relative magnitude NF deviation
\begin{align}
\varepsilon_{\text{m}} &= 20\,\text{log}_{10}\left(\dfrac{\norm{\left|\mat A \vec z\right|-\left|\vec b\right|}}{\norm{\vec b}}\right)\label{eq:NF}
\end{align}
can be evaluated.
For each ratio of the total number of measurements to unknowns, $\num{5000}$ random realizations of the forward operators, reference solution vectors and their corresponding measurement vectors were considered. For reasons of comparability, results for formulation~\eqref{eq:PC_LC_implementation} without the addition of the linear combination terms\,---\,thus without any partial coherence\,---\,have also been added.
It becomes evident from Fig.\,\ref{fig:synth_result} that the nonlinear formulation exploiting the knowledge of all available magnitudes and phase differences achieves better success rates for fewer measurement samples. This is expected to some extent, as the proposed formulation is only linear since $q-1$ restrictions on the magnitudes of the $\vec \psi$ terms are ignored\,---\,a loss of \enquote{information content}. However, once a certain ratio of measurements to unknowns is available, the linear formulation always returns the unique and correct solution. This ratio is equivalent to the bound derived in~\cite{Kornprobst.07.02.2020} as a necessary condition for convergence. Due to the differences in the formulations, this threshold in~\cite{Kornprobst.07.02.2020} can, however, not directly be proven to be valid for the approach in~\eqref{eq:linear_main}. The utilized nonlinear formulation for exploiting partially coherent observations may return a suboptimal solution even for large numbers of measurements\,---\,a success rate of $\num{100}\%$ is actually never achieved. This is due to the nonconvex nature of the formulation and can cause unreliable behavior in real-world applications. With neglect of the information about the partial coherence, i.e., ignoring that $\vec b_2$ and $\vec b_3$ have been acquired simultaneously with two coherent channels, the success rate drops further and stagnates at $\approx \num{95}\%$ in Fig.\,\ref{fig:synth_result}.

\setlength{\fw}{0.35\textwidth}
\setlength{\fh}{4cm}

\subsection{Synthetic Antenna Data}
\begin{figure}[t]
	\centering
	\includegraphics[width = 0.4\textwidth]{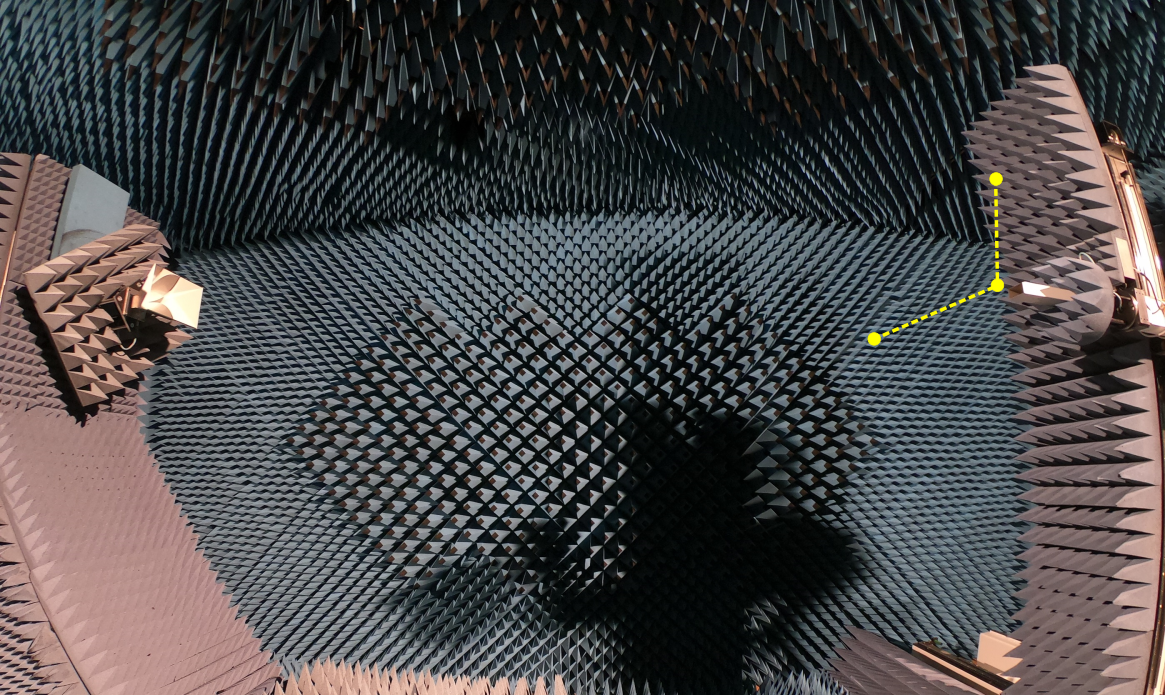}
	\caption{Near-field measurement arrangement of the DRH18 AUT (left) measured with an open-ended-waveguide probe (right). Three spherical measurements are performed sequentially, each with a different location of the probe on the planar positioner (yellow dots). The data is then combined as if the three measurements would have been performed in parallel, via a three-element \enquote{L}-shaped probe array and a receiver with three coherent channels. Here, we only consider synthetic data modeling this setup. }
	\label{fig:FFsetup}
\end{figure}%
As a final verification, spherical antenna NF data of a dual-ridged horn antenna (DRH18~\cite{drh18}) at $\SI{2.6}{\GHz}$ was processed. The real-world measurement data with full phase information was utilized to generate an equivalent AUT model via a vector spherical wave expansion~\cite{Hansen.1988}. With this model, NF data for a virtual spherical measurement with a probe antenna array consisting of three elements in an \enquote{L}-arrangement was generated. Within the transversal probe coordinate system, the elements are placed at the locations $\left(\SI{0}{\meter},\SI{0}{\meter}\right)$, $\left(\SI{1}{\meter},\SI{0}{\meter}\right)$ and $\left(\SI{0}{\meter},\SI{0.8}{\meter}\right)$, respectively, forming the three corners of an \enquote{L}. This virtual setup is illustrated in Fig.\,\ref{fig:FFsetup}, showing a possible realization in an anechoic chamber with an open-ended-waveguide probe. 

For modeling the AUT during the phaseless transformations, $n = \num{800}$ Hertzian and Fitzgerald dipoles placed tangentially on an enclosing box were used. Since different sources for the data generation and for the treatment of the inverse problem are used, inverse crime is avoided. 
From the virtual spherical measurements, $\num{2000}$ samples per set were randomly picked, resulting in a total number of $m = \num{6000}$ measurements available for the phase retrieval algorithms\,---\,plus the inherent phase differences between the probe antennas. Noise according to a signal-to-noise ratio (SNR) of $\SI{60}{\dB}$\,---\,with respect to the maximum value of the measurement signals\,---\,was added to the stacked measurement vector. The nonconvex solvers again started from an initial guess computed via the spectral method based on $\num{40}$ power iterations. 

\begin{figure}[t]
	\centering
		\subfloat[]{%
		\includegraphics{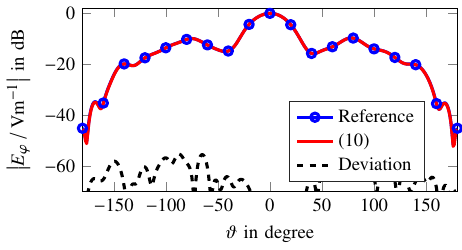}%
		\label{fig:FF_linear}%
	}%

	\subfloat[]{%
		\includegraphics{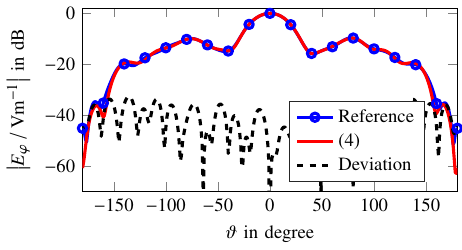}%
		\label{fig:FF_combs}%
	}%

	\subfloat[]{%
		\includegraphics{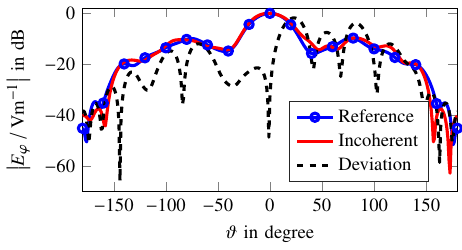}%
		\label{fig:FF_incoherent}%
	}%
	\caption{Dominant FF component of the electric field in the $\varphi = \SI{90}{\degree}$ cut plane of the DRH18 horn antenna at \SI{2.6}{\GHz} obtained with three phase retrieval formulations. Data was generated via an equivalent model of the DRH18 based on a spherical multipole expansion. The transformation results are normalized to their respective maximum values, at which their phase is also set to zero. a) Transformation exploiting partial coherence with~\eqref{eq:linear_main}. b) Transformation exploiting partial coherence with~\eqref{eq:PC_LC_implementation}. c) Transformation ignoring partial coherence in~\eqref{eq:PC_LC_implementation}. }
	\label{fig:FF}
\end{figure}

In Fig.\,\ref{fig:FF}, the obtained FF radiation characteristic of the dominant field component in the $\varphi = \SI{90}{\degree}$ cut plane through the main beam is depicted. The FF computed from the spherical multipole expansion was taken as reference. The deviation curves display the absolute value of the \textit{complex-valued difference} between the reference and the transformation result, normalized by the maximum value of the reference. While the graphs only display the magnitude of the electric field, large deviations can also be caused by a false phase distribution, as it seems to be the case in Fig.\,\ref{fig:FF}(c). In agreement with the relative NF deviations reported in Tab.\,\ref{tab:01}, the proposed linear transformation results in good FF accuracy, better than that of the nonlinear transformation exploiting local phase relations in~\eqref{eq:PC_LC_implementation}.  Ignoring the incomplete phase information, we obtain the inferior transformation result in Fig.\,\ref{fig:FF}(c). Essentially, the transformation result in Fig.\,\ref{fig:FF}(a) is limited by the SNR, while the results in Fig.\,\ref{fig:FF}(b) and~\ref{fig:FF}(c) are deteriorated due to local stationary points.
For completeness, the obtainable deviations between the spherical vector wave function representation and the dipole representation, when full phase information is available, is given in Tab.\,\ref{tab:01}, labeled as \enquote{coherent}.

\setlength{\fw}{0.35\textwidth}
\setlength{\fh}{3.1cm}

 \begin{table}
	\renewcommand{\arraystretch}{1.3}
	\caption{NF deviations obtained for the DRH18 horn antenna model}
	\label{tab:01}
	\centering
	\begin{tabular}{c |c c c | c}
		&  \eqref{eq:linear_main} &\eqref{eq:PC_LC_implementation} & Incoherent & Coherent \\
		\hline
		$\varepsilon_{\text{c}}$ &  $\SI{-50.9}{\dB}$ & $\SI{-30.2}{\dB}$ & $\SI{-3.2}{\dB}$ & $\SI{-55.6 }{\dB}$\\
		$\varepsilon_{\text{m}}$ &  $\SI{-53.0}{\dB}$ & $\SI{-40.0}{\dB}$ & $\SI{-33.7}{\dB}$ & $\SI{-58.6 }{\dB}$\\
	\end{tabular}
\end{table}

\section{Conclusion}
A reliable linear formulation for the task of phase retrieval with partially coherent measurements was presented and investigated. Suitable near-field data can be obtained by receivers with two or more coherent channels and appropriate probe arrays. The presented equations are highly flexible with respect to the form and availability of the knowledge of partial coherence and allow for a computationally efficient implementation. While effectively dropping a portion of the available magnitude restrictions, the algorithm features superior reliability compared to an existing nonconvex approach once a sufficient number of measurements has been acquired. The required number of samples is in the order of that needed for a conventional transformation with full phase information. Only in the regime of few available measurement samples, existing nonconvex approaches should be preferred as they do not drop information and may, by chance, return a more accurate result.

\bibliographystyle{IEEEtran}
\bibliography{IEEEabrv,EUCAP2021_phaseless}

\balance
\end{document}